\documentclass[superscriptaddress,pre,aps,twocolumn,showpacs,floatfix]{revtex4}

\usepackage[dvips]{graphicx}
\usepackage{amsmath}
\usepackage{amssymb}
\usepackage[nice]{nicefrac}
\usepackage{color}
\usepackage{amsmath}
\usepackage{grffile}
\usepackage{times}

\newcommand{\Fmax}{\mathcal{F}^{\mathrm{max}}_N}
\newcommand{\Fmin}{\mathcal{F}^{\mathrm{min}}_N}

\newcommand{\Smax}{{S}^{\mathrm{max}}_N}
\newcommand{\Smin}{{S}^{\mathrm{min}}_N}

\begin{document}
\bibliographystyle{prsty}
\title{Quantifying resonant activation like phenomenon in non-Markovian systems}

\author{Krzysztof Szczepaniec}
\email{kszczepaniec@th.if.uj.edu.pl}
\affiliation{Marian Smoluchowski Institute of Physics, and Mark Kac Center for Complex Systems Research, Jagiellonian University, ul. Reymonta 4, 30--059 Krak\'ow, Poland }

\author{Bart{\l}omiej Dybiec}
\email{bartek@th.if.uj.edu.pl}
\affiliation{Marian Smoluchowski Institute of Physics, and Mark Kac Center for Complex Systems Research, Jagiellonian University, ul. Reymonta 4, 30--059 Krak\'ow, Poland }

\date{\today}

\begin{abstract}
Resonant activation is an effect of a noise-induced escape over a modulated potential barrier.
The modulation of a energy landscape facilitates the escape kinetics and makes it optimal as measured by the mean first passage time. 
A canonical example of resonant activation is a Brownian particle moving in a time-dependent potential under action of Gaussian white noise.
Resonant activation is observed  not only  in typical Markovian-Gaussian systems but also in far from equilibrium and far from Markovianity regimes.
We demonstrate that using an alternative to the mean first passage time, robust measures of resonant activation, the signature of this effect can be observed  in general continuous time random walks in modulated potentials even in situations when the mean first passage time diverges.

\end{abstract}

\maketitle

\section{Introduction\label{sec:introduction}}

In the last decades, the concept of noise in physical systems has advanced from an unwanted addition to data,
towards subject of  interest of non-equilibrium statistical physics.
Noise is used to simplify  descriptions of dynamical systems when the detailed character of interactions is unknown or too complicated for exact methods. Noise is responsible for the occurrence of the so-called noise-induced effects. Among them stochastic resonance \cite{gammaitoni1998,anishchenko1999} and resonant activation \cite{doering1992} are the most popular. Usually, it is assumed that noise is white and Gaussian, however both non-Gaussian and non-Markovian extensions are also possible \cite{goychuk2003,goychuk2004}.
The examination of the constructive role of noises in physical systems has attracted considerable attention during the last two decades.

Resonant activation \cite{doering1992,boguna1998} is a resonant effect in which a noise-induced transition is further facilitated by a modulation of the energy landscape. A deterministic or stochastic modulation of the potential barrier can improve the system performance leading to the optimal escape kinetics as measured by the mean first passage time. Such a situation is observed for a white Gaussian noise \cite{doering1992}, $\alpha$-stable L\'evy type noises \cite{dybiec2007b,dybiec2004,dybiec2007,dybiec2009}, colored noise \cite{munakata1985,bag2003} or even in situations when the potential barrier is modulated by a colored process \cite{hanggi1994,iwaniszewski2003,novotny2000,majee2005}.

Here, we extend studies on resonant activation into a non-Markovian and non-Gaussian regime. We study a non-Markovian continuous time random walk scheme in a stochastically modulated potential when jump lengths are generated according to heavy-tailed distributions. The non-Markovianity of the studied process emerges due to slower-than-exponential decay of the waiting time distribution. This process is also non-Gaussian because jump lengths are determined by a heavy-tailed distribution with the diverging second moment.
On the one hand, the description of anomalous systems can be provided by
the continuous time random walk scheme which allows for a relatively easy treatment (exact or asymptotic) of various waiting time and jump length distributions resulting in the unified description of Markovian/non-Markovian and Gaussian/non-Gaussian systems \cite{metzler1999,metzler2000}.
On the other hand, such realms due to a slow decay of the waiting time distribution and power-law tails of the jump length distribution are described by the bi-fractional diffusion equation \cite{metzler1999,magdziarz2007b,magdziarz2007,Sokolov2006} which provides an equivalent description of the continuous time random walk framework.

As the waiting time distribution becomes of the power-law, heavy-tailed type, the system drifts into a non-Markovian regime.
Consequently, the standard way of identifying resonant activation fails, as the mean first passage time diverges due to heavy-tailed distributions of waiting times.
While this could be taken as an indication of the disappearance of the resonant activation phenomenon, we show that the effect of optimal escape kinetics can still be detected. In order to prove the existence of resonant activation, we use robust, quantile-based measures, which can be defined regardless of the existence of the mean first passage time. Quantile-based measures provide robust characteristics of the first passage time density and as a such can be used to expose the signature of resonant activation.
In this study, in addition to the examination of the properties of first passage time distributions, we also use extreme statistics and their properties. Extreme statistics provides additional insight into systems dynamics. In particular, it is  well suited for an examination of the system's performance and quantifying the stochastic resonance phenomenon.

The next section presents the model under consideration, defines measures of resonant activation and shows obtained results.
The paper is closed with summary and conclusions.

\section{Model and results\label{sec:model}}

The classical resonant activation setup \cite{doering1992} consists of a Brownian particle moving in a time-dependent potential field, which is switching dichotomously between two linear slopes of different heights. For the sake of simplicity, it is assumed that the dichotomous process is symmetric and Markovian \cite{horsthemke1984}, i.e. a potential stays in one of configurations for an exponentially distributed time. Additionally, the domain of motion is restricted to the finite interval with the reflecting boundary on the left and the absorbing boundary on the right. Here, we examine properties of the archetypal resonant activation setup when the Brownian particle is replaced by a more general random walker performing a general form of a continuous time random walk \cite{montroll1965,montroll1984,metzler2000}.

\begin{figure}[!ht]
%
%
%
%

\includegraphics[angle=0,width=0.99\columnwidth]{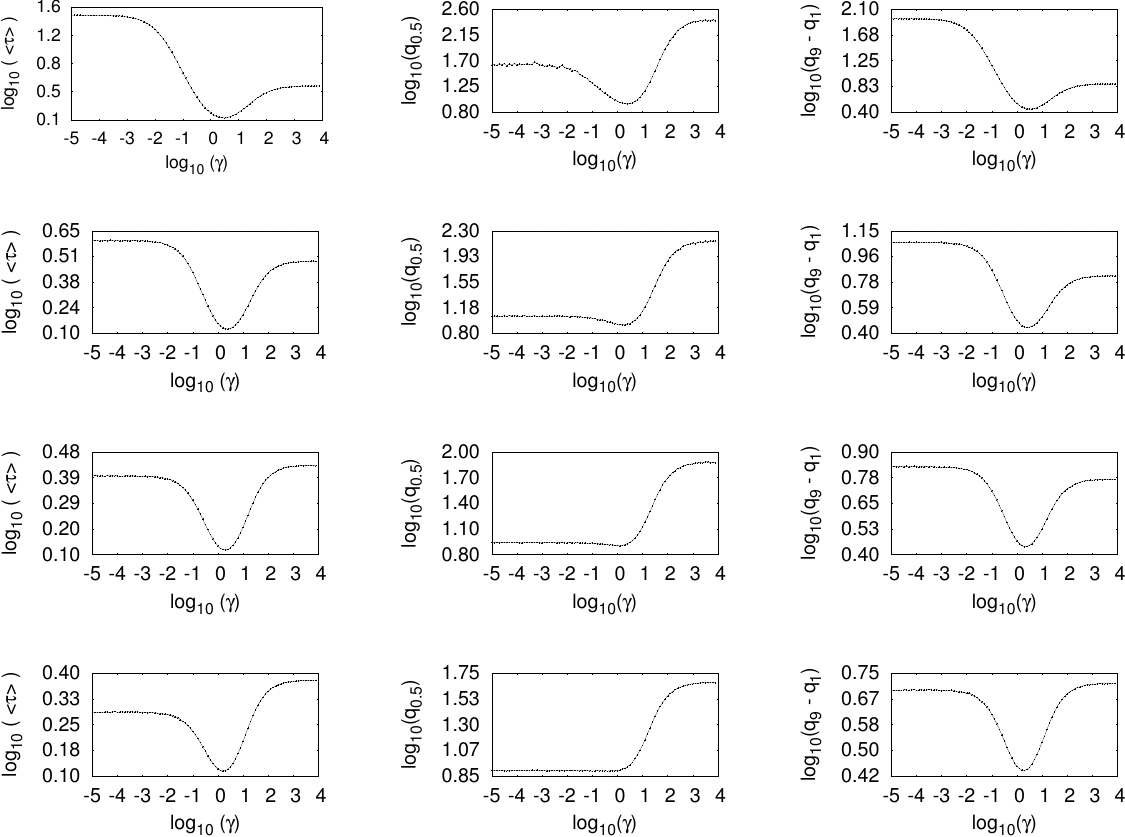} \\
\caption{Various measures of resonant activation: mean first passage time $\langle \tau \rangle$ (left column), median location $q_{0.5}$ (middle column) and inter-quantile width $q_{0.9}-q_{0.1}$ (right column) for the Markovian ($\nu=1$) case with $\alpha=\{2,1.9,1.8,1.7\}$ (from top to bottom).}
\label{fig:markovian}
\end{figure}

The subordination methods \cite{magdziarz2007b,magdziarz2008} are used in order to study a system described by the bi-fractional time dependent diffusion equation \cite{metzler1999,magdziarz2007b,magdziarz2007,Sokolov2006}:
\begin{equation}
 \frac{\partial p(x,t)}{\partial t}=\left[ \frac{\partial}{\partial x} V_{\pm}'(x,t) + \sigma^\alpha \frac{\partial^\alpha}{\partial |x|^\alpha} \right]{}_{0}D^{1-\nu}_{t} p(x,t).
\label{eq:ffpe}
\end{equation}
The bi-fractional Smoluchowski-Fokker-Planck equation describes the evolution of the probability density of finding a particle at the time $t$ in the vicinity of $x$. 
The subordination methods are based on the generation of a stochastic process whose evolution of the probability density is described by Eq.~(\ref{eq:ffpe}).
This is achieved by linking physical time $t$ with operational time $s$ by an appropriate time transformation.
The used method is briefly presented in the Appendix~\ref{sec:nummethod} while the detailed description can be found in \cite{magdziarz2007,magdziarz2007b,magdziarz2008,weron2008}.

In Eq.~(\ref{eq:ffpe}), ${}_{0}D^{1-\nu}_{t}$ denotes the Riemannn-Liouville fractional time derivative ${}_{0}D^{1-\nu}_{t}=\frac{\partial}{\partial t}{}_{0}D^{-\nu}_{t}$ defined by the relation
\begin{equation}
{}_{0}D^{1-\nu}_{t}f(x,t)=\frac{1}{\Gamma(\nu)}\frac{\partial}{\partial t}\int^{t}_0 dt'\frac{f(x,t')}{(t-t')^{1-\nu}}
\end{equation}
and $\frac{\partial^{\alpha}}{\partial |x|{\alpha}}$ stands for the Riesz-Weil fractional space derivative with the Fourier transform
\begin{equation}
{\cal{F}}\left[\frac{\partial^{\alpha}}{\partial |x|^{\alpha}} f(x)\right]=-|k|^{\alpha}\hat{f}(x).
\end{equation}
The potential $V_\pm(x,t)$ dichotomously switches between two linear configurations characterized by two distinct heights $H_\pm$, i.e. $V_\pm(x,t)=H_\pm x$. Initially, a test particle is located at the origin and the configuration of the potential is set to $V_+(x,t)$ or $V_-(x,t)$ with equal probabilities. As in the Doering Gadoua model \cite{doering1992} the dichotomous process is symmetric and Markovian. Consequently, it is described by a single parameter $\gamma$ which is the rate of the potential switching, i.e. a dichotomous process takes two possible values only and stays constant for the exponentially distributed time. The domain of motion is restricted to the finite interval, i.e. at $x=0$ there is a reflecting boundary, while at $x=1$ there is an absorbing boundary. Boundaries impose additional constraints on the probability density $p(x,t)$ which are hard to implement on the operator level \cite{Buldyrev2001,Buldyrev2001a,zoia2007} but are relatively easily controlled on the single trajectory 
level. Eq.
~(\ref{eq:ffpe}) extends the classical resonant activation model into a non-equilibrium
non-Markovian regime.

\begin{figure}[!ht]
%
%
%
\includegraphics[angle=0,width=0.99\columnwidth]{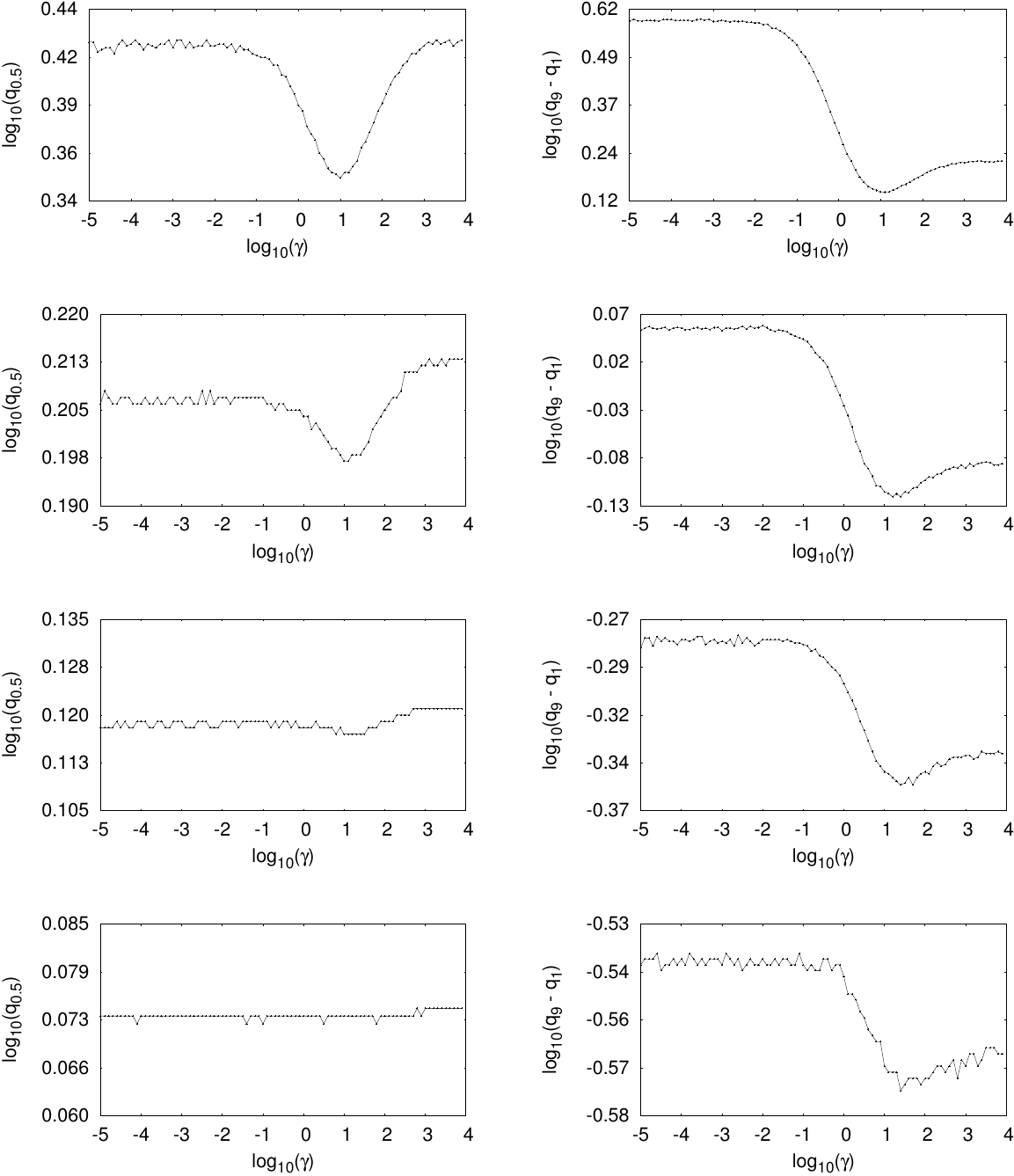} \\
\caption{Various measures of resonant activation: median location $q_{0.5}$ (left column) and inter-quantile width $q_{0.9}-q_{0.1}$ (right column) for the non-Markovian case with $\nu=0.9$ and $\alpha=\{2,1.9,1.8,1.7\}$ (from top to bottom).}
\label{fig:nu09}
\end{figure}

\begin{figure}[!ht]
%
%
%
\includegraphics[angle=0,width=0.99\columnwidth]{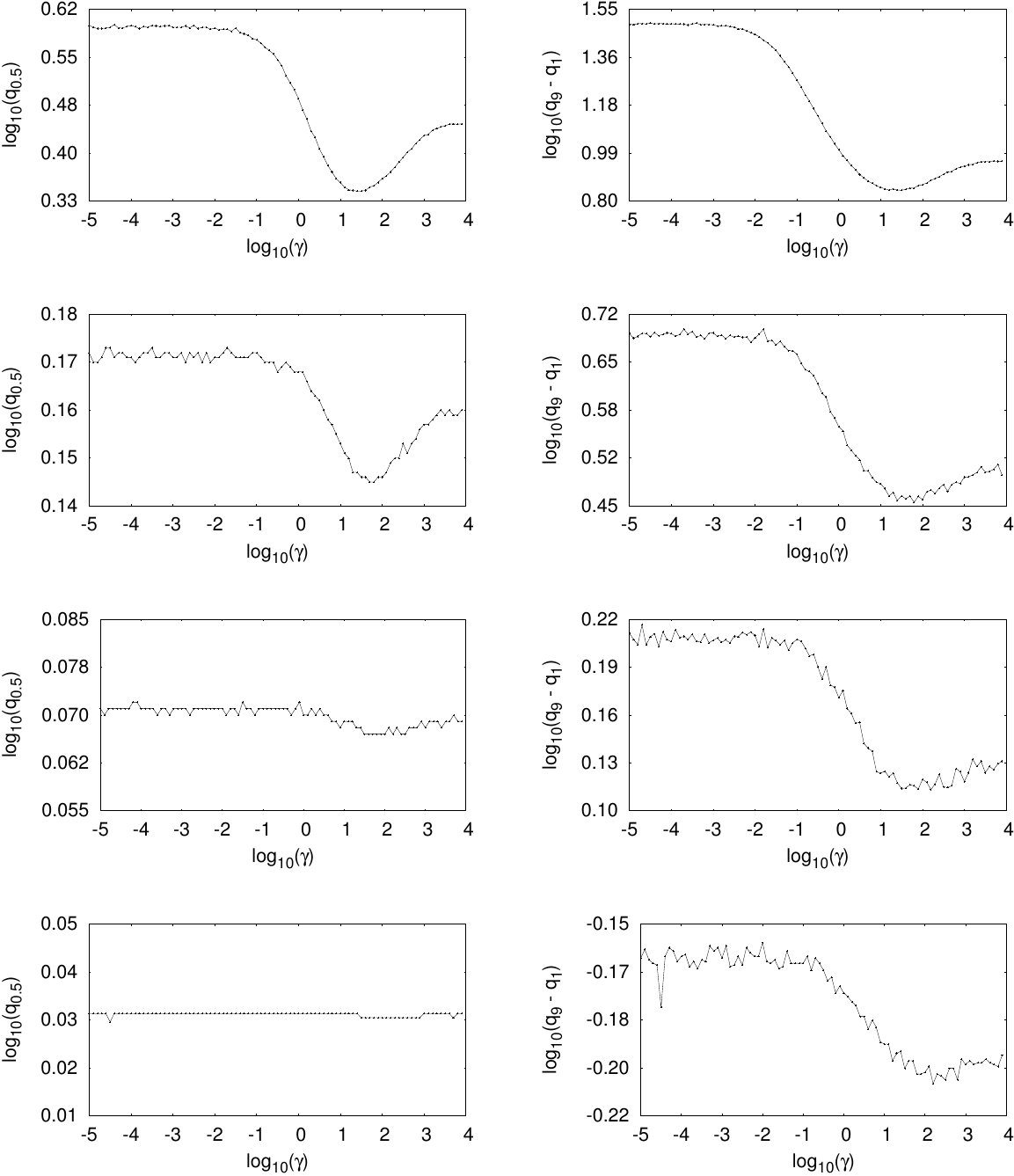} \\
\caption{Various measures of resonant activation: median location $q_{0.5}$ (left column) and inter-quantile width $q_{0.9}-q_{0.1}$ (right column) for the non-Markovian case with $\nu=0.6$ and $\alpha=\{2,1.9,1.8,1.7\}$ (from top to bottom).}
\label{fig:nu06}
\end{figure}

From the microscopic point of view, the model describes an escape kinetics of a random walker from the finite interval restricted by absorbing and reflecting boundaries.
In the force free case, Eq.~(\ref{eq:ffpe}) emerges as a description of a continuous time random walk scenario with power law distributed waiting times, $p(\Delta t) \propto \Delta t^{-(\nu+1)}$ and jump lengths, $p(\Delta x) \propto |\Delta x|^{-(\alpha+2)}$, with $0 < \alpha < 2$ and $0 < \nu <1$. Here the jump length distribution is also modified by the external potential, which slightly alters the jump length distribution. Nevertheless, its asymptotics is still of the power-law type. Such a continuous time random walk scenario is both non-Markovian and non-Gaussian \cite{dybiec2009h}. The Markovian Gaussian scenario is recovered for $\nu \geqslant 1$ with $\alpha \geqslant 2$. The subdiffusion parameter $\nu$ controls the level of non-Markovianity while $\alpha$ controls the level of non-Gaussianity.
Eq.~(\ref{eq:ffpe}) is a general type equation which is also obtained when jump lengths are distributed according to symmetric $\alpha$-stable densities \cite{feller1968,nolan2002,yanovsky2000,schertzer2001}, which for $\alpha<2$ have required $|\Delta x|^{-(\alpha+1)}$ asymptotics.  We refer to $\nu$ as the subdiffusion parameter and to $\alpha$ as the stability index. The studied random walk is characterized by the first passage time $\tau$
\begin{equation}
 \tau = \min\{t>0 \;\;:\;\; x(0)=0 \mbox{ and }  x(t) \geqslant 1 \}.
\end{equation}
From the set of the first passages times $\tau_i$ it is possible to calculate various characteristics of the given system which can be used to measure the performance of escape kinetics.

Traditionally, the strength of resonant activation is measured by the mean first passage time, but such a quantifier can be calculated only when the first passage time density is not of the heavy-tailed type.
In particular, for $\nu=1$, the seminal resonant activation setup is recovered. The first passage time density has exponential tails and it has a well-defined mean value.
Otherwise, the mean first passage time diverges and cannot be used to characterize the performance of escape kinetics.
This is the case with the studied model in the non-Markovian regime, i.e. for $\nu<1$, when the first passage time density has power-law tails
with the exponent $-(\nu+1)$.
Nevertheless, that is the point which requires special attention.
The calculation of resonant activation characteristics is based on a large, yet finite sample of first passage times.
The finite size effects result in the effective truncation of the first passage time density.
Consequently, from a finite sample it is still possible to calculate the mean first passage time.
However, due to heavy-tails, this quantity fluctuates very strongly and becomes meaningless.

For a general type of continuous time random walks which can be characterized by the diverging waiting time different measures have to be employed.
Therefore, one needs to consider more robust measures which can be based on distributions of the first passage time or directly derived from this density, e.g. quantiles of the first passage time distribution
\begin{equation}
\mathrm{Prob}(\tau \leqslant q_p)=p\;\;\;\;\; (0<p<1),
\end{equation}
i.e. such a value of the first passage time that the probability to find a smaller value than $q_p$ is $p$. Quantiles can be also expressed by the cumulative density of the first passage time $\mathcal{F}(t)$ by the relation $\mathcal{F}(q_p)=p$.
In particular, one can use the median location ($q_{0.5}$) or the inter-quantile width ($q_{0.9}-q_{0.1}$) in order to measure the system's performance. The efficiency of quantile-based measures relies on the fact that optimal escape kinetics should affect the properties of the first passage time distributions what should be further manifested by the dependence of quantiles on control parameters (in particular the switching rate $\gamma$).

\begin{figure}[!ht]
\includegraphics[angle=0,width=0.99\columnwidth]{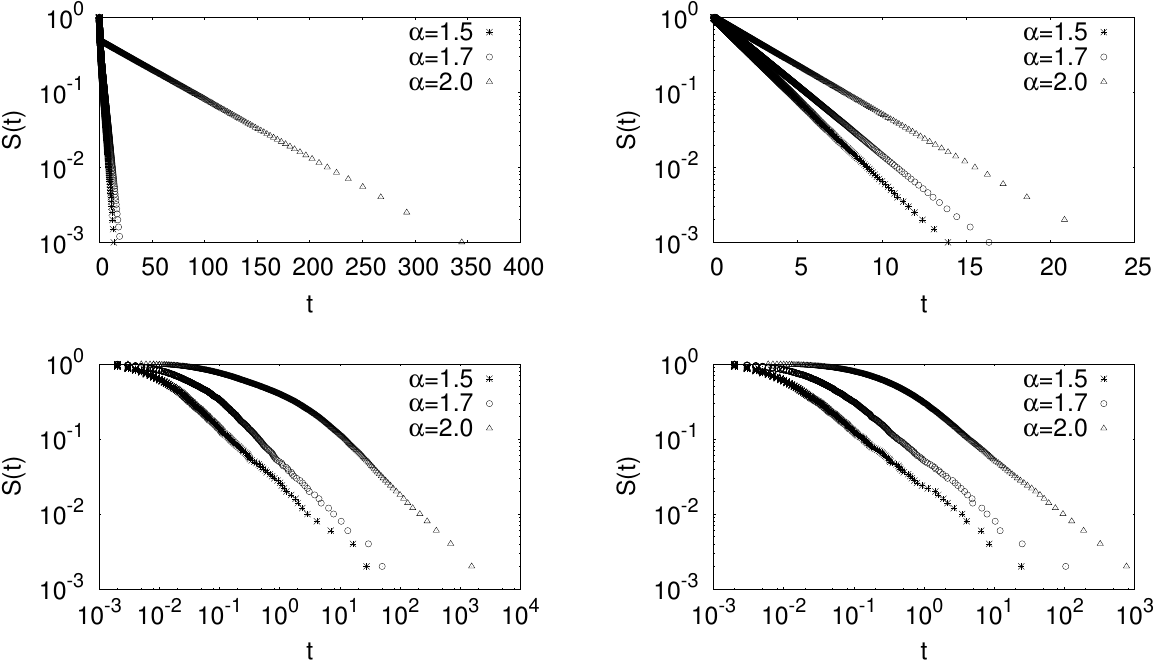} \\
 \caption{Survival probability $S(t)=1-\mathcal{F}(t)$ for low switching rate $\gamma$ (left column) and high switching rate $\gamma$ (right column) for $\nu=1$ (top row) and $\nu=0.7$ (bottom row). Various lines correspond to various $\alpha=\{2.0,1.7,1.5\}$.
Please note log-lin (top row) and log-log (bottom row) scales.
 }
 \label{fig:fptd}
\end{figure}

\begin{figure}[!ht]

\includegraphics[angle=0,width=0.99\columnwidth]{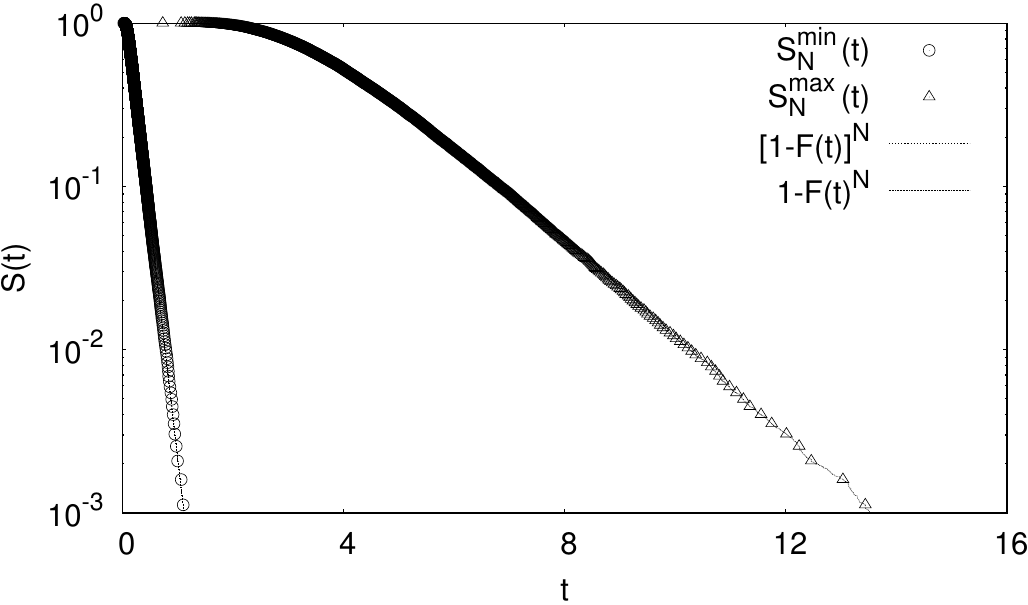} \\
\caption{Survival probability $S(t)=1-\mathcal{F}(t)$ for extreme statistics: 
$\Smin(t)$ and $\Smax(t)$. Solid lines present theoretical results obtained from 
``single particle simulations'' while points represent results of ``$N=10$'' 
particles simulation. Other simulation parameters: switching rate 
$\gamma=10$, 
number of repetitions $N_{\mathrm{rep}}=10^6$, time step of integration $\Delta 
t=10^{-3}$, subdiffusion paremeter $\nu=1$ and stability index $\alpha=2$.}
\label{fig:extremecdf}
\end{figure}

Within the model properties of the resonant activation phenomenon are studied in the non-Gaussian ($\alpha<2$) and non-Markovian regime ($\nu<1$). However, for the reference point, also the Markovian and the Gaussian case is studied ($\nu=1$ with $\alpha \leqslant 2$). It is assumed that the potential switches between $V_\pm(x,t)=H_\pm x$ configurations with $H_+=8$ and $H_-=0$. Changes in the height of the potential barrier $H_\pm$ are described by a symmetric Markovian dichotomous process with the rate $\gamma$, i.e. the dichotomous process stays constant for an exponentially distributed time with the intensity equal to the switching rate $\gamma$, see \cite{horsthemke1984,doering1992}.

In order to study properties of alternative measures of resonant activation, the Markovian non-Gaussian case ($\nu=1$ with $\alpha \leqslant 2$) was studied. Fig.~\ref{fig:markovian} shows the dependence of the mean first passage time $\langle \tau \rangle$, the median location $q_{0.5}$ and the inter-quantile width $q_{0.9}-q_{0.1}$ on $\gamma$ for relatively large values of the stability index only, i.e. $\alpha=\{2.0,1.9,1.8,1.7\}$ from top to bottom. In such a situation the first passage time distribution has a well-defined mean and consequently the phenomenon of resonant activation can be captured by the examination of the mean first passage time, see left column of Fig.~\ref{fig:markovian}.
In line  with earlier studies, strong deviations from the Gaussian distribution of jump lengths result in diminishing  resonant activation, see Fig.~\ref{fig:markovian} and \cite{dybiec2004}. However, a full disappearance of the effect is observed for lower values of the exponent $\alpha$ than presented in Fig.~\ref{fig:markovian}, see \cite{dybiec2004}.

Resonant activation is understood as an optimal escape kinetics as measured by the mean first passage time. Decrease in the mean first passage time is a consequence of changes in the first passage time distribution and corresponds to the narrowing of the first passage time density. This can be measured not only by the mean first passage time but also by the width of the first passage time density which can be characterized by the inter-quantile distance, e.g. $q_{0.9}-q_{0.1}$. The inter-quantile distance provides a robust measure, because it exists independently of the existence of the mean or the variance of the first passage time distribution. The inter-quantile distance is sensitive to the modulation of the barrier and properly captures resonant activation, as it can be observed in the right column of Fig.~\ref{fig:markovian}. Nevertheless, the narrowing of the first passage time distribution is not necessarily reflected in the dependence of the median location
of the first passage time density on the rate of the potential switching $\gamma$, see the middle column of Fig.~\ref{fig:markovian}. Therefore, the median can be used as a measure for the phenomenon of resonant activation only for a very limited set of system parameters.

\begin{figure}[!ht]
%
%
%
%
\includegraphics[angle=0,width=0.99\columnwidth]{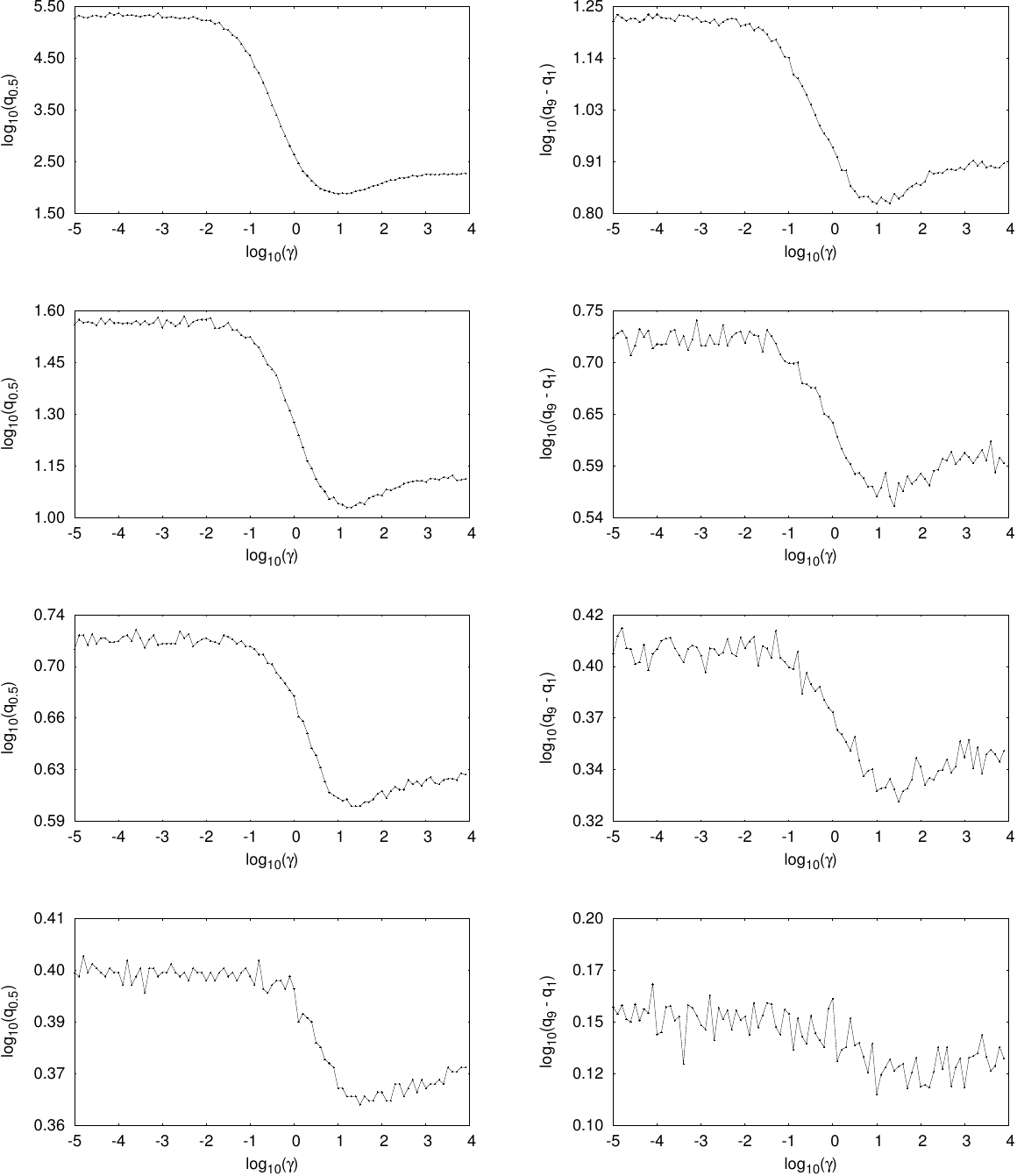} \\
\caption{Various measures of resonant activation derived from the order statistics $\Fmax(t)$ with $N=10$: median location $q_{0.5}$ (left column) and inter-quantile width $q_{0.9}-q_{0.1}$ (right column) for the non-Markovian case with $\nu=0.9$ and $\alpha=\{2,1.9,1.8,1.7\}$ (from top to bottom).}
\label{fig:extremenu09}
\end{figure}

\begin{figure}[!ht]
%
%
%
\includegraphics[angle=0,width=0.99\columnwidth]{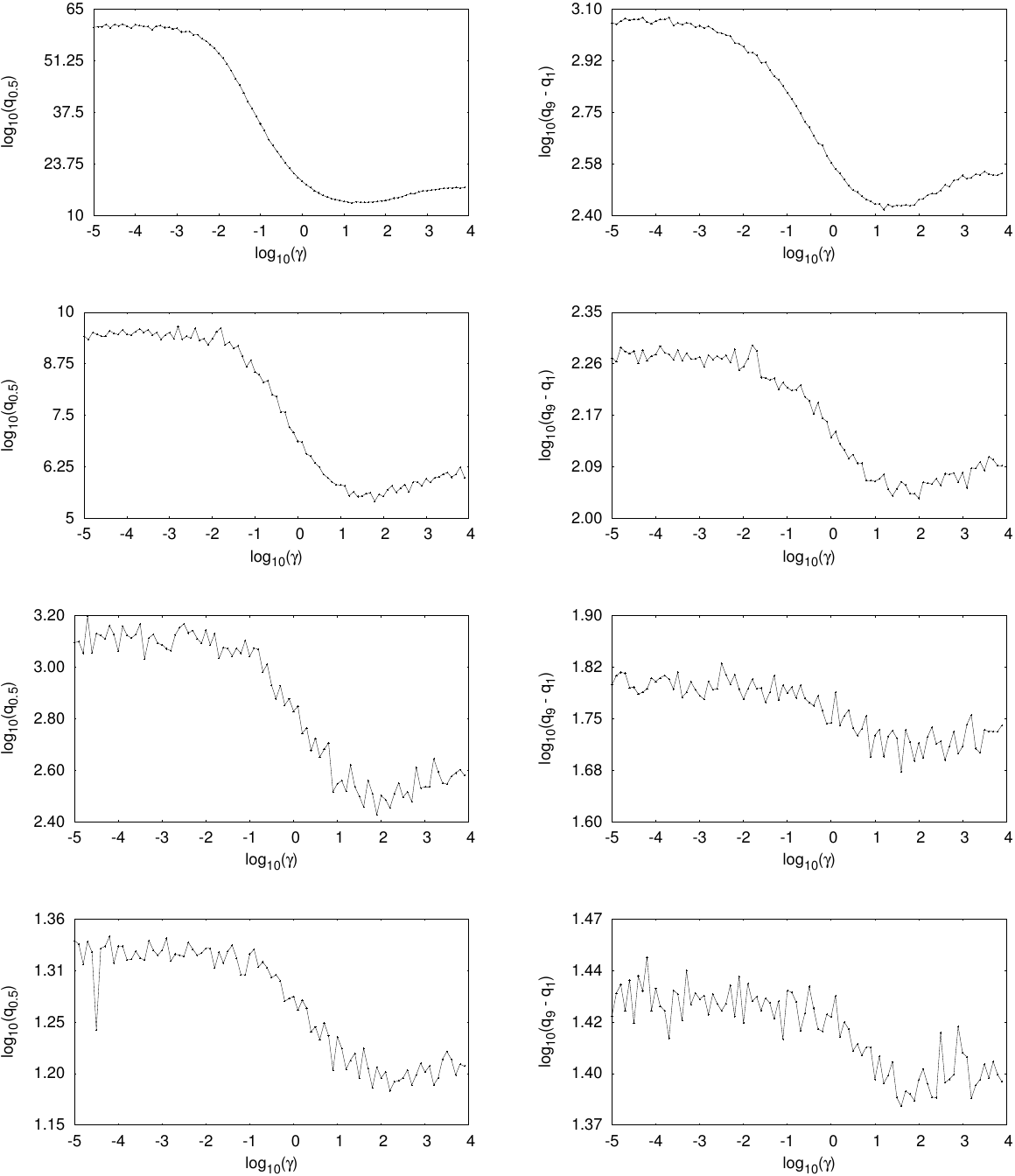} \\
\caption{Various measures of resonant activation derived from the order statistics $\Fmax(t)$ with $N=10$: median location $q_{0.5}$ (left column) and inter-quantile width $q_{0.9}-q_{0.1}$ (right column) for the non-Markovian case with $\nu=0.6$ and $\alpha=\{2,1.9,1.8,1.7\}$ (from top to bottom).}
\label{fig:extremenu06}
\end{figure}

Both possible initial states of the potential are equally probable. Thus, in the Markovian case, for a low and moderate switching rate $\gamma$ a half of the events correspond to fast escapes (over a low barrier configuration) while the other half corresponds to slow escapes (high barrier configurations). This makes the median not very sensitive as a measure of resonant activation. The different situation is with the inter-quantile width which is of the same monotonicity as the mean first passage time, because it neglects some fraction of the fastest and slowest escape events.

In the non-Markovian case ($\nu<1$), the mean first passage time, due to the divergence of the mean waiting time for a next jump, diverges and cannot be used to characterize the resonant activation effect. Therefore, measures which can be used in such cases are quantile-based. Consequently, the median and the inter-quantile width will be used to characterize the system performance. Moreover, for $\nu<1$ the system has only one characteristic time scale, i.e. the one associated with the barrier switching process which is $1/\gamma$.

Figures~\ref{fig:nu09} and~\ref{fig:nu06} demonstrate quantile-based measures of resonant activation for a selected set of non-Markovian systems. For $\nu<1$ the inter-quantile width clearly demonstrates that the dependence of the width of the first passage time density is a non-monotonous function of the switching rate $\gamma$. One can find an optimal switching rate leading to the lowest width of the first passage time density. The non-monotonous dependence of the width of the first passage time distribution is a signature of a phenomenon similar to resonant activation. With decreasing the value of the jump length exponent $\alpha$ the resonant activation effect disappears, compare various rows of Figs.~\ref{fig:nu09} and~\ref{fig:nu06}. Likewise, the decrease in the waiting time exponent $\nu$ weakens the
effect, compare Fig.~\ref{fig:nu09} and Fig.~\ref{fig:nu06}.

Figure~\ref{fig:fptd} shows the survival probability of a random walker in various scenarios including a Gaussian-Markovian case. For the low switching rate $\gamma$ (with $\alpha=2$ and $\nu=1$), the survival probability  attains a pronounced double exponential form, see the top left panel of Fig.~\ref{fig:fptd}. As the jump length distribution departs from the Gaussian, a random walker can perform longer jumps with a significantly larger probability. This facilitates the escape kinetics which could be corroborated by the smaller average time required to reach the absorbing boundary. This is further confirmed by the decrease of the survival probability with the decrease of the stability index $\alpha$. On the one hand, when the system drifts into the non-Markovian regime, subdiffusion becomes stronger, escape kinetics is hindered and long waiting times cause the survival probability to be non-negligible for a significantly longer time. On the other hand, it is well visible that longer jumps result in lower
values of the survival probability indicating an increase in the efficiency of the escape kinetics with the decrease of the stability index $\alpha$, compare the top and bottom rows of Fig.~\ref{fig:fptd}.

For $\nu<1$,  the median behaves in a more informative way than in the Markovian case, but quickly looses sensitivity to resonant activation as the non-Gaussianity of the system progresses.The dependence of the median and the inter-quantile width on the switching rate $\gamma$ clearly confirms the occurrence of the resonant activation effect. While both the median and the inter-quantile width can be used as indicators of resonant activation, in a strongly non-Gaussian and non-Markovian regime both undergo fluctuations, rendering both measures less reliable for determining the exact position of an optimal switching rate, see the bottom rows of Figs.~\ref{fig:nu09} and~\ref{fig:nu06}.

Measures characterizing resonant activation do not need to be based on single particle simulations. Instead of considering a single particle system it is possible to study the motion of $N$ independent (non-interacting) particles. The presence of multiple random walkers allows to calculate minimal and maximal first passage times. The minimal first passage time is the exit time of the fastest random walker out of $N$, while the maximal exit time is the exit time of the last (slowest) random walker out of $N$. If random walkers are independent, instead of examining  $N$ random walkers it is also possible to study a single random walker and to divide data into $N$ elements disjoint sets. For every set the minimal (maximal) first passage time is just a minimal (maximal) exit time of an $N$ random walker system. The properties of order statistics are well known \cite{david1970,sefling1980}. In particular, it is known that first order statistics has a significantly faster decay than the probability density of an
original variable. While the $N$-th (last) order statistics has the same asymptotics as the original variables, i.e. the first passage times $\tau$, see \cite{dybiec2010e}.

The first (minimum) order statistics accounts for the fastest events only. It does not provide sufficient resolution to measure the system performance and as a such is not very sensitive to the manipulation of the model parameters (results not shown). The maximum statistics is suitable for quantifying the resonant activation effect because it uses only maximal first passage times. The width of this distribution is more sensitive to the manipulation of the model parameters. Consequently, the same analysis which was performed in Figs.~\ref{fig:markovian} -- \ref{fig:nu06} is conducted for the maximum (last order) statistics, see Figs.~\ref{fig:extremenu09} and~\ref{fig:extremenu06}.

Figure~\ref{fig:extremecdf} demonstrates that maximal and minimal first passage time densities constructed from the simulation ($N=10$) are exactly the same as the ones estimated from the single-particle experiment. More precisely, the single-particle experiment allows to estimate $\mathcal{F}(t)$ which in turn can be used to calculate $\Fmax(t)=[\mathcal{F}(t)]^N$ and $\Fmin(t)=1-[1-\mathcal{F}(t)]^N$, see \cite{david1970,sefling1980}. Instead of $\mathcal{F}(t)$, $\Fmax(t)$ and $\Fmin(t)$ it is more convenient to inspect the survival probability $S(t)=1-\mathcal{F}(t)$, because it reveals asymptotic behavior more clearly. For illustrative purposes, $\alpha=2$ with $\nu=1$ have been chosen.

\begin{figure}[!ht]
\includegraphics[angle=0,width=0.5\columnwidth]{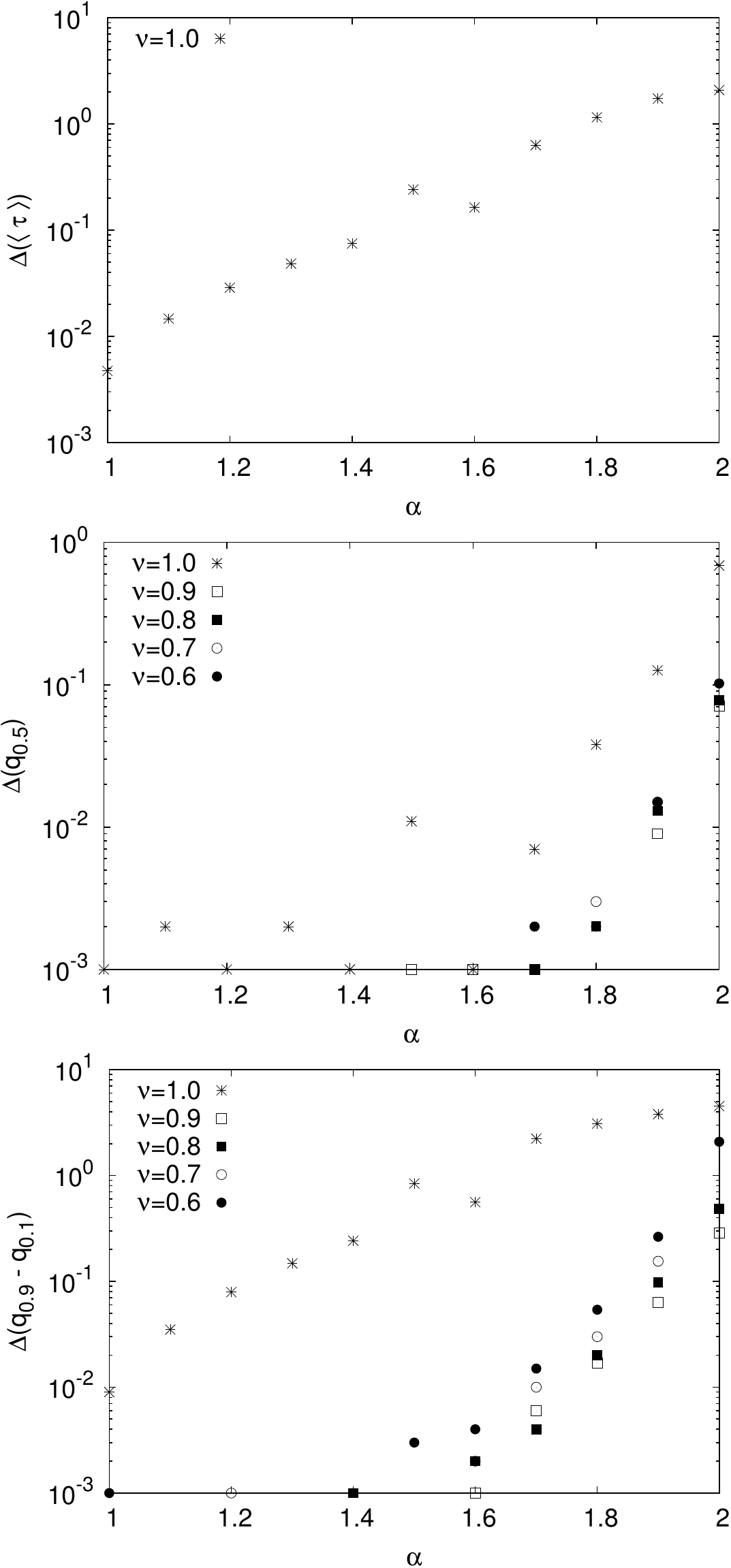} \\
\caption{Strength of resonant activation as measured by the relative depth of: mean first passage time $\langle \tau \rangle$ (top panel), median location $q_{0.5}$ (middle panel) and inter-quantile width $q_{0.9}-q_{0.1}$ (bottom panel). Various curves correspond to various values of the subdiffusion parameter $\nu$.}
\label{fig:delta}
\end{figure}

\begin{figure}[!ht]
\includegraphics[angle=0,width=0.5\columnwidth]{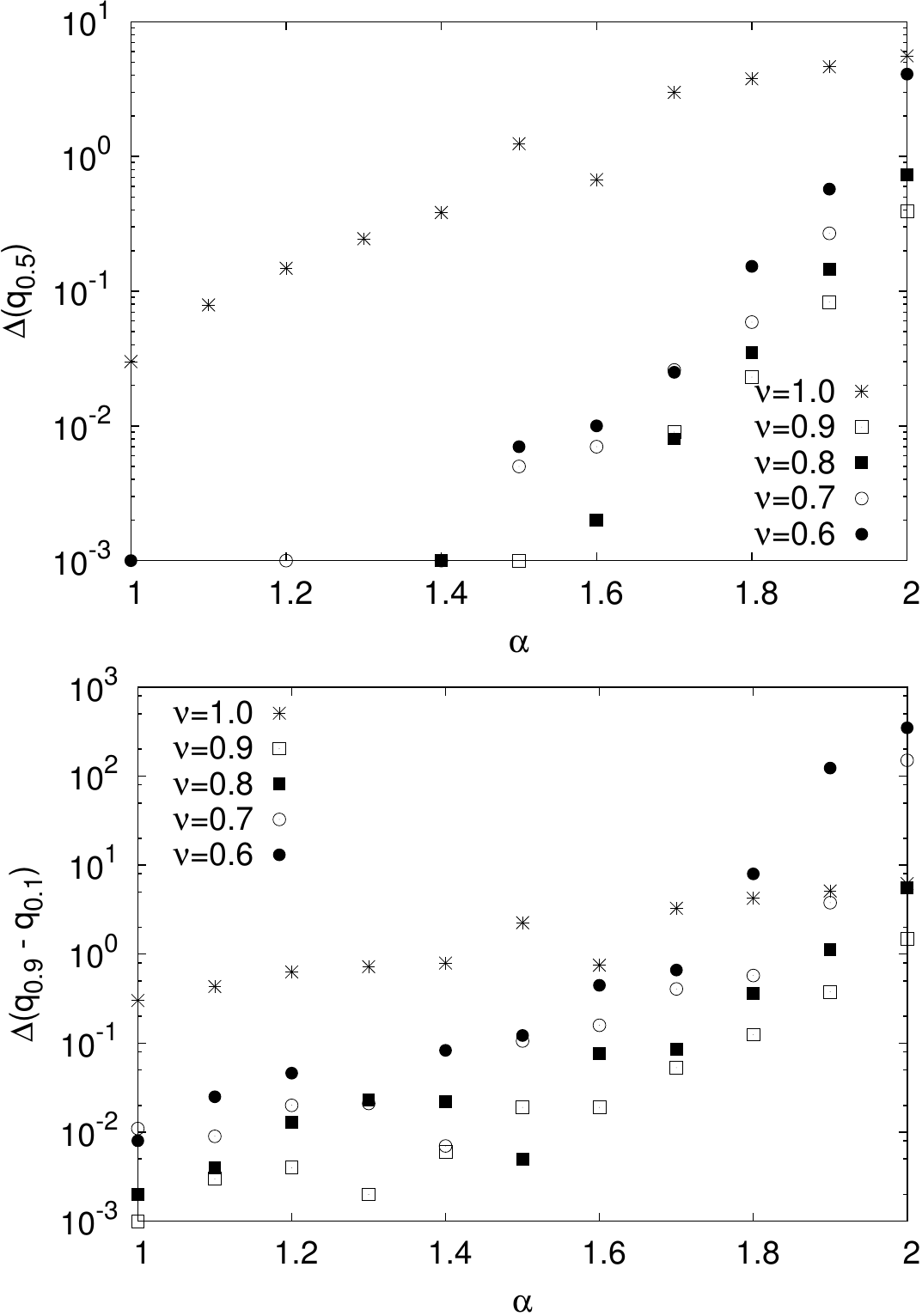} \\
\caption{Strength of resonant activation as measured by the relative depth of extreme statistics: median location $q_{0.5}$ (top panel) and inter-quantile width $q_{0.9}-q_{0.1}$ (bottom panel). Various curves correspond to various values of the subdiffusion parameter $\nu$.}
\label{fig:extremedelta}
\end{figure}

Figures~\ref{fig:extremenu09} and~\ref{fig:extremenu06} show the dependence of the median of the maximum (last order) statistics for a $N=10$ particle system. Due to smaller statistics, fluctuations are larger. Among two measures of the escape kinetics efficiency (median and inter-quantile width) the median position is more robust with respect to fluctuations (in the single-particle simulations it is the opposite). Nevertheless, the analysis of the last order statistics supports the main conclusion that with the decrease of the stability index $\alpha$ and the subdiffusion parameter $\nu$ resonant activation disappears.

Figures~\ref{fig:markovian} -- \ref{fig:nu06} and~\ref{fig:extremenu09} -- \ref{fig:extremenu06} show quantifiers as a function of the switching rate $\gamma$. In order to determine the existence of the resonant activation effect, the strength of the effect is measured by the deviation of the minimal value of a given quantifier from the lowest asymptotics of that quantifier, i.e.
\begin{equation}
 \Delta(u)=\min\{u(\gamma=-\infty),u(\gamma=\infty)\}-\min(u),
\end{equation}
where $u$ could be the mean first passage time ($\langle \tau \rangle$), the median ($q_{0.5}$) or the inter-quantile width ($q_{0.9}-q_{0.1}$). Such a quantifier measures the separation of the minimum of a given characteristics from its lower asymptotics, which can be reached either for a low or large switching rate $\gamma$. Therefore, it is the relative depth of the minimum (if it exists).

Figures~\ref{fig:delta} and~\ref{fig:extremedelta} show the separation of the minima of resonant activation measures from lower asymptotics ($\Delta(\dots)$) for all quantifiers used: the median and the inter-quantile width. The values of $\Delta$ for any of quantifiers used with $\nu<1$ are in general smaller than for $\nu=1$ indicating the weakening of the resonant activation effect when the system departs from the Markovian regime. Analogously, resonant activation weakens with decreasing the value of the jump length exponent $\alpha$, i.e. when the jump length distribution becomes heavy-tailed.
The situation is very similar for extreme statistics when the range of the relative depth variation is larger than in the single-particle case, compare Fig.~\ref{fig:delta} and Fig.~\ref{fig:extremedelta}.
The changes in the jump length distribution ($\alpha$) and waiting time distribution ($\nu$) modify the system performance resulting in the disappearance of resonant activation with the decrease of $\alpha$ and $\nu$ exponents as measured by $\Delta$.

\section{Summary and Conclusions\label{sec:summary}}

Resonant activation is one of classical effects demonstrating a constructive role of noises in physical systems. It proves that the noise-induced escape events can be further optimized by an additional modulation of the energy landscape. Typically, resonant activation has been studied in Markovian systems and it is considered as a generic property of a barrier crossing process in a modulated energy landscape. Here, we show that resonant activation is not only observed in a plenitude of noise-driven systems but its signature is also visible in non-Markovian and non-Gaussian systems. However, in such realms, it has to be inspected by more robust measures than the typically used mean first passage time (which can diverge).

The classical Gaussian-Markovian  case was used as a test bench in order to verify whether quantile-based measures can be used to analyze properties of resonant activation. This test confirmed suitability of such measures. Next, quantile-based measures were applied in the non-Markovian regime in order to prove existence of the effect similar to resonant activation also in situations when, due to long trapping events, the mean first passage time diverges. Such a signature of resonant activation is visible in general continuous time random walks in time modulated potentials. The strength of the effect weakens with the widening of the jump length distribution and waiting time distribution, finally, resulting in the disappearance of resonant activation.


\appendix
\section{Numerical methods\label{sec:nummethod}}
The applied numerical scheme is based on the subordination method \cite{piryatinska2005,eule2009,feller1968} which has been extended \cite{magdziarz2007,magdziarz2007b,magdziarz2008,weron2008} to give a proper stochastic representation of
trajectories of the process $X(t)$ whose evolution of the probability density is described by the bi-fractional Smoluchowski-Fokker-Planck equation \cite{metzler1999,magdziarz2007b,magdziarz2007,Sokolov2006}
\begin{equation}
 \frac{\partial p(x,t)}{\partial t}=\left[ \frac{\partial}{\partial x} V'(x,t) + \sigma^\alpha \frac{\partial^\alpha}{\partial |x|^\alpha} \right]{}_{0}D^{1-\nu}_{t} p(x,t).
\label{eq:ffpea}
\end{equation}

The solution $p(x,t)$ of Eq.~(\ref{eq:ffpea}) can be estimated from the probability density of the subordinated process
\begin{equation}
 X(t)=\tilde{X}(S_\nu(t))
\end{equation}
where $\tilde{X}(s)$ fulfills the following Langevin equation
\begin{equation}
 d \tilde{X}(s) = -V'(\tilde{X}(s),U(s))ds + \sigma dL_{\alpha,0}(s)
\end{equation}
driven by the standard $\alpha$-stable motion \cite{janicki1994}.
The inverse-time subordinator $S_\nu(t)$ is defined
\begin{equation}
 S_\nu(t)=\mbox{inf}\{\tau: U(s) > t  \},
\end{equation}
where $U(s)$ stands for a strictly increasing $\nu$-stable L\'evy motion
whose Laplace transform is $\langle \exp(-kU(s)) \rangle = \exp(-\tau k^\nu)$ \cite{janicki1994}.
The inverse $\nu$-stable subordinator links the real time $t$ with the operational time $s$.

In the general case of the time dependent force $F(x,t)= -\frac{\partial}{\partial x}V(x,t) $, displacements due to the deterministic force acting on a particle are approximated as $-V'(\tilde{X}(s),U(s))ds$, see \cite{magdziarz2008,weron2008}. This assures that the force acting on a particle is changing in the physical time $t$, see \cite{magdziarz2008,weron2008}.
In the less general case of the time independent force, the deterministic force is approximated by $-V'(\tilde{X}(s))$ and the fractional Smoluchowski-Fokker-Planck equation can be rewritten in the usual form, i.e. with the Riemannn-Liouville fractional time derivative acting on the whole right hand side of Eq.~(\ref{eq:ffpea}).
Approximation of the stochastic processes $X(s)$ and $U(s)$ requires generation of pseudo random numbers distributed according to the symmetric $\alpha$-stable densities ($X(s)$) or the totally skewed ($\nu<1$) $\nu$-stable densities ($U(s)$) which can be generated according to well known formulas \cite{chambers1976,weron1996}.
For the detailed  description of the subordination method and its sample implementation see original papers \cite{magdziarz2007,magdziarz2007b,magdziarz2008,weron2008}.


\begin{acknowledgments}
Computer simulations have been performed at the Academic
Computer Center Cyfronet AGH (Krak\'ow, Poland) under CPU grant
MNiSW/Zeus\_lokalnie/UJ/052/2012.

\end{acknowledgments}

\end{document}